\documentclass{ws-ijmpcs}

\usepackage{graphicx}
\usepackage{epsfig}
\usepackage{mathrsfs,bm,overpic}

\newcommand{\CP}{\ensuremath{\mathit{CP}}}
\newcommand{\Acp}{\ensuremath{A_{\CP}}}
\newcommand{\acp}[1]{\ensuremath{a_{\CP}^{\text{#1}}}}
\newcommand{\Acpraw}{\ensuremath{\Acp^{\text{raw}}}}
\newcommand{\stat}{\ensuremath{\mathit{~(stat.)}}}
\newcommand{\syst}{\ensuremath{\mathit{~(syst.)}}}
\newcommand{\Dbar}{\ensuremath{\overline{D}{}}}

\begin{document}

\markboth{Michael J. Morello}
%{Instructions for Typing Manuscripts (Paper's Title)}
{\CP\ violation in the $D^0\to\pi^+\pi^-$ decay at CDF}

\catchline{}{}{}{}{}
%
%%%%%%%%%%%%%%%%%%%%%%%%%%%%%%%%%%%%%%%%%%%%%%%%%%%%%%%%%%%%%%%%%%%%

%\title{MEASUREMENT OF THE \CP\ ASYMMETRY IN THE $D^0\to\pi^+\pi^-$ DECAY AT CDF
\title{\CP\ VIOLATION IN THE $D^0\to\pi^+\pi^-$ DECAY AT CDF
%INSTRUCTIONS FOR TYPESETTING 
%MANUSCRIPTS
%\footnote{For the title, try not to use more than 
%3 lines. Typeset the title in 10 pt roman, uppercase and 
%boldface.}
}

\author{MICHAEL JOSEPH MORELLO\footnote{On behalf of the CDF Collaboration}}

\address{Fermi National Accelerator Laboratory, Pine St and Kirk Rd,\\
Batavia, Illinois 60510, USA \\
morello@fnal.gov}

\maketitle

\begin{history}
\received{Day Month Year}
\revised{Day Month Year}
\end{history}

\begin{abstract}

We report a measurement of the \CP\ violating asymmetry in \mbox{$D^0\to\pi^+\pi^-$} decays 
using approximately 215,000 decays reconstructed in about $5.94$ fb$^{-1}$ of CDF data.
 We use the strong $D^{\star +}\to D^0\pi^+$ decay  (``$D^{\star}$ tag'') to identify the 
flavor of the charmed meson at production time and exploit \CP-conserving strong $c\bar{c}$ 
pair-production in $p\bar{p}$ collisions. Higher statistic samples of Cabibbo-favored 
$D^0\to K^-\pi^+$ decays with and without $D^{\star}$ tag are used to highly suppress 
systematic uncertainties due to detector effects. The result, 
\mbox{$\Acp(D^0 \to \pi^{+}\pi^{-}) = \bigl[0.22\pm0.24\stat\pm0.11\syst\bigr]\%$}, is the world's most 
precise measurement to date and it is fully consistent with no \CP\ violation.

\end{abstract}

\ccode{PACS numbers: 14.40.Lb, 13.25.Ft }

\section{Introduction and motivation}

Time integrated \CP-violating asymmetries of singly-Cabibbo transitions 
as $D^0\to\pi^+\pi^-$ and $D^0\to K^+ K^-$ are powerful probes of new physics (NP). 
Contribution to these decays from ``penguin'' amplitudes are negligible in the 
Standard Model (SM), but presence of NP particles could enhance the size of 
\CP-violation with respect to the SM expectation. Any asymmetry significantly 
larger than few $0.1\%$, as expected in the CKM hierarchy, may unambiguously 
indicate new physics contributions \cite{theory}. We present a measurement of 
time-integrated \CP\ violating asymmetry in the Cabibbo-suppressed $D^0\to\pi^+\pi^-$ decay:
\begin{equation}\label{eq:acp}
\Acp(D^0\to\pi^+\pi^-)=\frac{\Gamma(D^0\to\pi^+\pi^-)-\Gamma(\Dbar^0\to\pi^-\pi^+)}{\Gamma(D^0\to\pi^+\pi^-)+\Gamma(\Dbar^0\to\pi^-\pi^+)}.
\end{equation}
The measured integrated asymmetry, owing to the slow mixing rate of 
charm mesons, reduces at first order to a sum of two terms:
\begin{equation}\label{eq:acp3}
\Acp(D^0 \to \pi^+\pi^-) = \acp{dir}+\int_0^\infty\Acp(t)\text{D}(t)dt\approx\acp{dir} + \frac{\langle t \rangle}{\tau}\acp{ind}
\end{equation}
where $t/\tau$ is the proper decay time in unit of $D^0$ lifetime ($\tau \approx 0.5$ ps). 
The first term arises from direct and the second one from mixing-induced CP violation. 
The integration in eq.~\eqref{eq:acp3} is performed over the observed distribution of proper 
decay time, $D(t)$. Since the value of $\langle t \rangle$ depends strongly on $D(t)$, different 
values of $\Acp$ could be observed in different experimental environments because of different 
sensitivities to $\acp{dir}$ or $\acp{ind}$. Since the trigger used in this analysis imposes 
requirements on minimum impact parameters of the $D^0$ decay particles, our sample is enriched 
of higher-valued proper decay time candidates with respect to B-factory experiments. 
This makes this analysis more sensitive to mixing-induced CP violation.

\section{Detector and trigger}
The CDF~II detector \cite{cdf} is a magnetic spectrometer surrounded by calorimeters 
and muon detectors. It provides a determination of the decay point of particles with 
15 $\mu$m resolution in the transverse plane using six layers of double-sided silicon-microstrip 
sensors at radii between 2.5 and 22 cm from the beam. A $96$-layer drift chamber extending 
radially from 40 to 140 cm from the beam provides excellent momentum resolution, yielding 
approximately 8  MeV/c$^2$ mass resolution for two body charm decays. A three-level trigger 
system selects events enriched in decays of long-lived particles by exploiting the presence 
of displaced tracks in the event and measuring their impact parameter with offline-like 
30 $\mu$m resolution. The trigger requires presence of two charged particles with transverse 
momenta greater than $2$~GeV/c, impact parameters greater than $100$ microns and basic cuts 
on azimuthal separation and scalar sum of momenta.

%\section{Analysis overview}
\section{Measurement}

We updated and improved an early Run II analysis \cite{giagu}, using an event sample collected 
with the trigger on impact parameter from March 2001 to January 2010 that corresponds at 
about $5.94$~fb$^{-1}$ of integrated luminosity.

We measure the asymmetry in singly-Cabibbo suppressed $D^0\to\pi^+\pi^-$ decays from $D^{*}$ through 
fits of the $D^0\pi^+$  and $\Dbar^0\pi^-$ distributions. The observed asymmetry include a possible 
tiny contribution from actual CP violation, diluted in much larger effects from instrumental 
charge-asymmetries. Indeed the layout of the main tracker detector, the drift chamber, is intrinsically
charge asymmetric due to a $35^\circ$ tilt angle of the cells from the radial direction \cite{cdf}, 
thus different detection efficiencies for positive and negative low-momentum tracks induce an instrumental 
asymmetry in the number of reconstructed $D^\star$-tagged $D^0$ and $\Dbar^0$ mesons. Other possible 
asymmetries may originate in slightly different performance of pattern-reconstruction and track-fitting 
algorithms for negative and positive particles. The combined effect of these is a net asymmetry in the 
range of a few percents. %, as shown in fig.~\ref{fig:soft}. 
This must be corrected to better than one per mil to match the expected statistical precision of the present measurement.

We exploit a fully data-driven method that uses higher statistic samples of $D^\star$-tagged 
(indicated with an asterisk) and untagged Cabibbo-favored $D^0\to K^-\pi^+$ decays to correct 
for all detector effects thus suppressing systematic uncertainties to below the statistical ones. 
The uncorrected ``raw'' asymmetries %\footnote{``Raw'' are the observed asymmetries in signal yields, 
%$$\Acpraw(D^0\to f) = \frac{N_{\text{obs}}(D^0\to f)-N_{\text{obs}}(\Dbar^0\to\bar{f})}{N_{\text{obs}}(D^0\to f)+N_{\text{obs}}(\Dbar^0\to\bar{f})},$$before any correction for instrumental effects has been applied.} 
in the three samples can be written as a sum of several (assumed small) contributions:
\begin{equation}\label{eq:acpraw}
\begin{aligned}
\Acpraw(\pi\pi^\star) &= \Acp(D^0 \to \pi^{+}\pi^{-}) + \delta(\pi_s)^{\pi\pi^\star}\\
\Acpraw(K\pi^\star) &= \Acp(D^0 \to K^{-}\pi^{+}) + \delta(\pi_s)^{K\pi^\star} + \delta(K\pi)^{K\pi^\star}\\
\Acpraw(K\pi) &= \Acp(D^{0} \to K^{-}\pi^{+}) + \delta(K\pi)^{K\pi},
\end{aligned}
\end{equation}
where  
$\Acp(D^0 \to \pi^{+}\pi^{-})$ and  $\Acp(D^0 \to K^{-}\pi^{+})$ are the actual physical asymmetries; 
$\delta(\pi_s)^{\pi\pi^\star}$ and $\delta(\pi_s)^{K\pi^\star}$ are the instrumental asymmetries in reconstructing a 
positive or negative soft pion associated respectively to a $\pi^+\pi^-$ and a $K^{-}\pi^{+}$ charm decay. 
$\delta(K\pi)^{K\pi}$ and $\delta(K\pi)^{K\pi^\star}$ are the instrumental asymmetries in reconstructing a $K^+\pi^-$ or a $K^-\pi^+$ charm decay respectively for the untagged and the $D^{*+}$--tagged case. 
The physical asymmetry is extracted by subtracting the instrumental effects through the combination~\cite{cdfnote:10296}:
\begin{equation}\label{eq:formula}
\Acp(\pi\pi) = \Acpraw(\pi\pi^\star) - \Acpraw(K\pi^\star) + \Acpraw(K\pi).
\end{equation}
We reconstruct approximately 215,000 $D^\star$--tagged $D^0\to\pi^+\pi^-$ decays, $5\times10^{6}$  $D^\star$--tagged $D^0\to\pi^+K^-$ 
decays and $29\times10^{6}$ $D^0\to\pi^+K^-$ decays where no tag was required. The much larger statistics of $D^0\to\pi^+K^-$ channels, 
used for correction of instrumental asymmetries, with respect to the signal sample ensures smaller systematic uncertainties than 
statistical ones on the final result. 
We extract independent signal yields  for $D^0$ and $\Dbar^0$ candidates without using particle identification in the analysis.
 In the two $D^\star$-tagged samples this is done using the charge of the soft pion. In the untagged $D^0\to K^-\pi^+$ sample we 
randomly divided the sample in two indipendent subsamples similar in size. In each subsample we calculate the mass of each 
candidate with a specific mass assignments: $K^-\pi^+$ in the first subsample and $K^+\pi^-$ in the second one. Thus in one sample 
the $D^0\to K^-\pi^+$ signal is correctly reconstructed and appears as a narrow peak, overlapping a broader peak of the misreconstructed 
$\Dbar^0\to K^+\pi^-$ component. The viceversa applies  the other sample. The raw asymmetry is extracted by fitting the number 
of candidates populating the two narrow peaks.

We determine the yields by performing a binned fit to the  $D^0\pi_s$-mass ($K\pi$-mass) distribution combining positive and negative 
decays of both tagged (untagged) samples. 
The resulting raw asymmetries are:
$\Acp^{\text{raw}}(\pi\pi^\star)= (-1.86\pm0.23)\%$,
$\Acp^{\text{raw}}(K\pi^\star)= (-2.91\pm0.05)\%$,
$\Acp^{\text{raw}}(K\pi) = (-0.83\pm0.03)\%$.

%\section{Systematic uncertainties}
The analysis technique has been extensively tested on Monte Carlo simulation using samples simulated with a 
wide range of physical and detector asymmetries to verify that the cancellation works regardless of the 
specific configuration. These studies confirm the validity of our approach and provide a quantitative 
estimate of possible asymmetries induced by higher order detector effect that may not get fully cancelled 
or effects of not factorization of $K\pi$ and $\pi_s$ reconstruction efficiencies. This upper limit is 
used as systematic uncertainty and amount to $0.009\%$.

We evaluate all other systematic uncertainties from data. In most cases, this implied varying slightly the 
shape of the functional forms used in fits, repeating the fit on data, and using the difference between 
the results of these and the central fit as a systematic uncertainty. This overestimate the size of the 
systematic effects because it introduces an additional statistical source of fluctuation in the results. 
But we can comfortably afford that given the large event samples size involved.
The dominant contributions to the systematic uncertainties on the asymmetry measurement come from 
the uncertainty on the differences in charge of the mass shapes, and the uncertainty due to the
contamination by charm mesons produced in $b$--hadron decays (CP--violating asymmetries in $B$ decays 
induce an asymmetric source of charm and anti-charm mesons).
We obtain a total systematic uncertainty on our final $\Acp(\pi\pi)$ measurement of $0.11\%$, approximately half 
of the statistical uncertainty.

\section{Final result and conclusions}
We report the measurement of the CP asymmetry in the decay $D^0\to\pi^+\pi^-$ using 5.94 fb$^{-1}$ of data collected by the CDF displaced track trigger. The final result is $$\Acp(D^0\to\pi^+\pi^-) = \bigl[+0.22\pm0.24\stat\pm0.11\syst\bigr]\%,$$ which is consistent with CP conservation and also with the SM predictions. 

To disentangle the independent contributions of direct and indirect CP violation in \mbox{$D^0\to\pi^+\pi^-$} decays, an analysis where the time evolution of charm decays is studied is needed. Nevertheless some interesting conclusions could be derived either comparing our result with B-factories measurements or making some theoretical assumptions.

\begin{figure}[t]
\centering
\begin{overpic}[width=0.45\textwidth,grid=false]{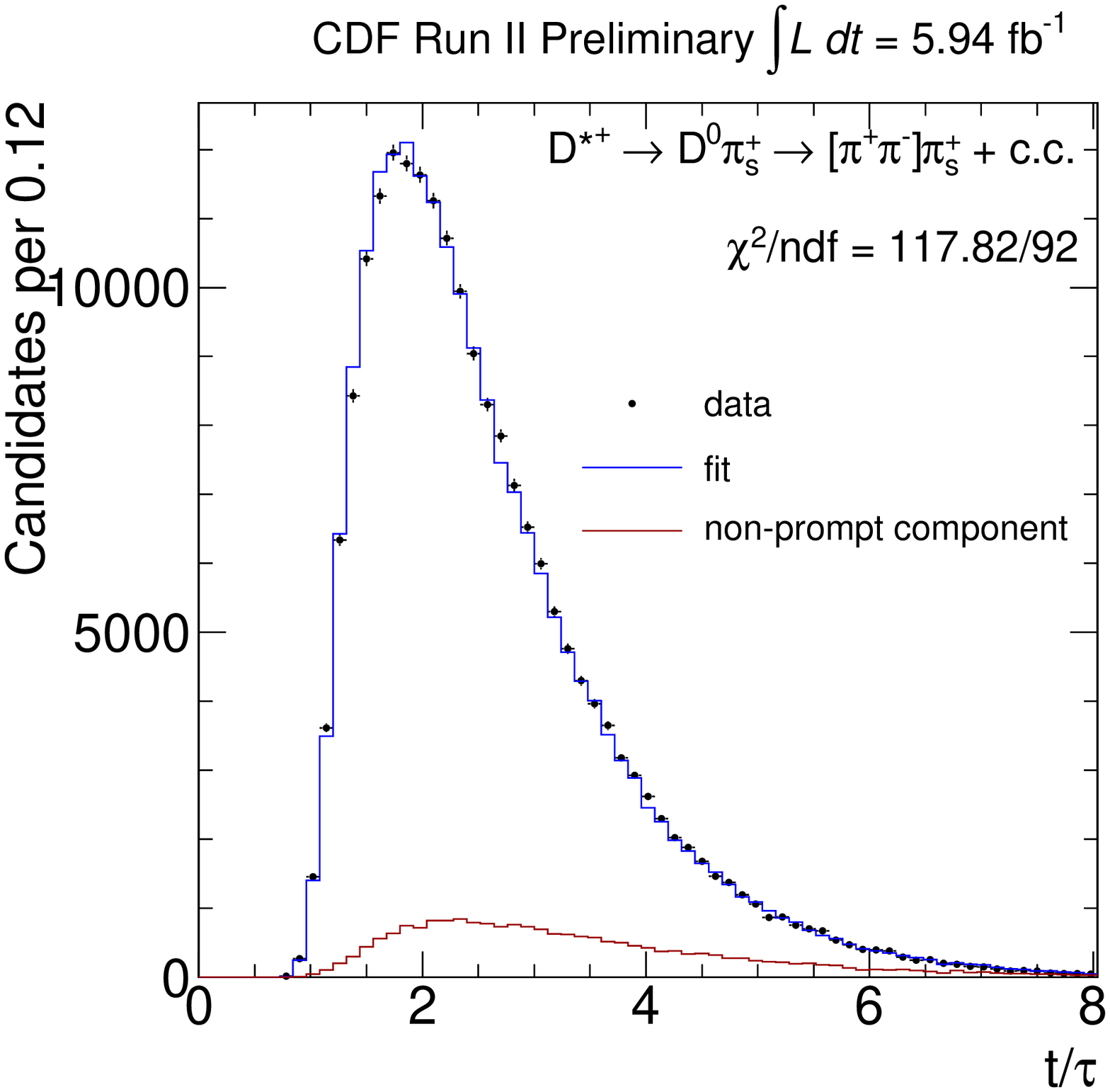}
%\begin{overpic}[width=0.45\textwidth,trim=10pt 168pt 20pt 0pt,clip,grid=false]{./figs/D0pipi_D0_t_fit.eps}
\put(21,78){(a)} 
\end{overpic}\hfil
\begin{overpic}[width=0.45\textwidth,grid=false]{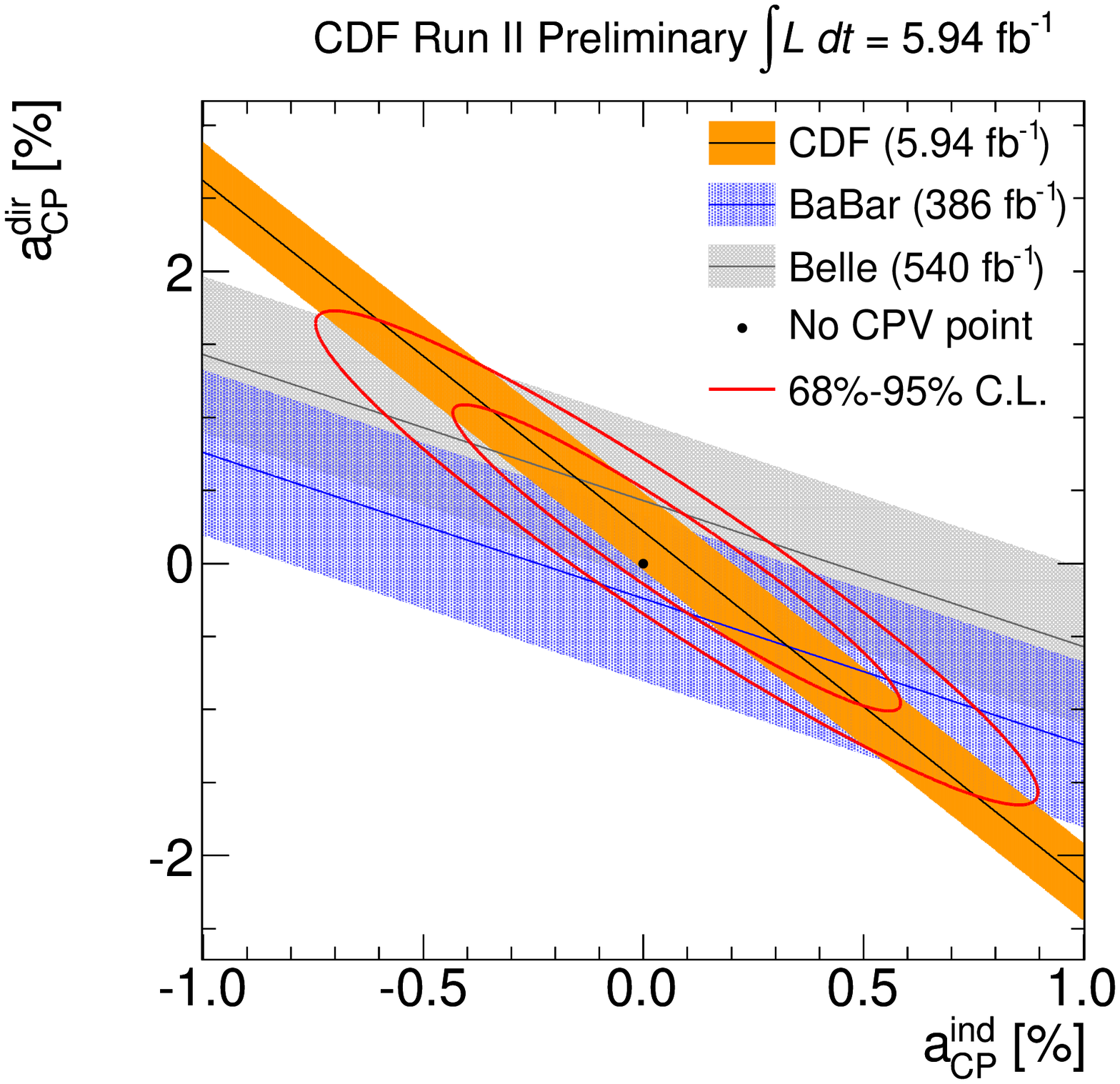}
\put(21,16){(b)}
\end{overpic}
\caption{Fit to the proper decay time (in units of $D^0$ lifetime) distribution of sideband-subtracted tagged $D^0\to\pi\pi$ data (a). Comparison of our measurement with current best results from B-factories in the parameter space $(\acp{ind},\acp{dir})$ (b).}\label{fig:combination}
\end{figure}

The observed asymmetry is at first order the linear combination of a direct, $\acp{dir}$, and an indirect, $\acp{ind}$, CP violating asymmetry through a coefficient that is the mean proper decay time of $D^0$ candidates in the data sample (see  eq.~\eqref{eq:acp3},). Fig.~\ref{fig:combination}~(a) shows a fit to the mean proper decay time distribution of our tagged \mbox{$D^0\to\pi^+\pi^-$} sample, the resulting mean value is $2.40\pm0.03\ (\mathit{stat.}+\mathit{syst.})$ times the $D^0$ lifetime. Our measurement therefore describes a straight band in the plane $(\acp{ind},\acp{dir})$ with angular coefficient $-2.4$. The same holds for B-factories measurements, with angular coefficient $-1$ \cite{babelle}, due to their unbiased acceptance in charm decay time. The three measurements in the plane $(\acp{ind},\acp{dir})$ are shown in fig.~\ref{fig:combination}~(b), where the bands are $1\sigma$ wide and the red curves represent the $68\%$ and $95\%$ CL limits of the combined result assuming Gaussian uncertainties.

If we assume no direct CP violation in the charm sector eq.~\eqref{eq:acp3} simplifies to
$$\Acp(\pi^+\pi^-) \approx \frac{\langle t \rangle}{\tau}\acp{ind}$$
so this measurement implies
$$\acp{ind} = \bigl[+0.09\pm0.10\stat\pm0.05\syst\bigr]\%,$$
that means the range $[-0.124,0.307]\%$ covers $\acp{ind}$ at the 95\% CL. Note that, since $\langle t \rangle/\tau$ in our sample is greater than in B-factories ones, this range is more than five times tighter than the ones obtained using B-factories measurements. %, as shown in fig.~\ref{fig:direct_and_indirect}~(a).

Conversely, assuming $\acp{ind}=0$, our number is directly comparable to other measurements in different 
experimental configurations. In this case, %fig.~\ref{fig:direct_and_indirect}~(b), 
our statistical uncertainties are half those from the best B-factories measurements, 
and also systematic uncertainties are smaller.

\section*{Acknowledgments}

Many thanks to the organizers of {\it Charm 2010} for their successful work.

%\begin{thebibliography}{000} %for 3 digits
%\begin{thebibliography}{00}  %for 2 digits

\end{document}